\documentclass{llncs}
\usepackage{amsmath}
\usepackage{amssymb}
\usepackage{enumerate}
\usepackage{longtable}

\newcommand{\boxi}{\ensuremath{\mathrm{box}}}
\newcommand{\cubi}{\ensuremath{\mathrm{cub}}}

\newcommand{\ignore}[1]{}

\newcommand{\surl}[1]{{\small\url{#1}}}

\title  {Representing graphs as the intersection of axis-parallel cubes}
\subtitle{\small{(Extended Abstract)}}

\author{L. Sunil Chandran \and Mathew C. Francis  
\thanks{Indian Institute of Science,
Dept. of Computer Science and Automation,
Bangalore--560 012, India.  email: \emph{mathew,sunil@csa.iisc.ernet.in}}
\and Naveen Sivadasan
\thanks{Advanced Technology Centre, TCS, Deccan Park, Madhapur,
Hyderabad--500 081, India. email: \emph{s.naveen@atc.tcs.com}}
\institute{}}

\begin{document}
\maketitle
\pagestyle{plain}
\pagenumbering{arabic}

\begin{abstract}
A unit cube in $k$ dimensional space (or \emph{$k$-cube} in short) is defined as
the Cartesian product $R_1\times R_2\times\cdots\times R_k$ where $R_i$(for 
$1\leq i\leq k$)
is a closed interval of the form $[a_i,a_i+1]$ on the real line.
A $k$-cube representation of a graph $G$ is a mapping of the vertices of $G$
to $k$-cubes such that two vertices in $G$ are adjacent if and only if
their corresponding $k$-cubes have a non-empty intersection.
The \emph{cubicity} of $G$, denoted as $\cubi(G)$, is the minimum $k$
such that $G$ has a $k$-cube
representation. Roberts \cite{Roberts} showed that for any graph $G$ on $n$
vertices, $\cubi(G)\leq 2n/3$. Many NP-complete graph problems
have polynomial time deterministic algorithms or have good approximation
ratios in graphs of low cubicity.
In most of these algorithms, computing a low dimensional cube representation
of the given graph is usually the first step.

From a geometric embedding point of view, a  $k$-cube representation of
$G=(V,E)$ yields an embedding $f: V \rightarrow \mathbb{R}^k$
such that for any two vertices $u$ and $v$, $||f(u) - f(v)||_{\infty} \le
1$ if and only if $(u,v) \in E$.

We present an efficient algorithm to compute the $k$-cube representation
of $G$ with maximum degree $\Delta$ in $O(\Delta\ln n)$ dimensions. We then
further strengthen this bound by giving an algorithm that produces a $k$-cube
representation of a given graph $G$ with maximum degree $\Delta$ in
$O(\Delta \ln b)$ dimensions where $b$ is the bandwidth of $G$. Bandwidth of
$G$ is at most $n$ and can be much lower. The algorithm takes as input a
bandwidth ordering of the vertices in $G$. Though computing the bandwidth
ordering of vertices for a graph is NP-hard, there are heuristics that
perform very well in practice. Even
theoretically, there is an $O(\log^4 n)$ approximation algorithm for
computing the bandwidth ordering of a graph using which our algorithm
can produce a $k$-cube representation of any given graph in $k=O(\Delta(\ln b
+ \ln\ln n))$ dimensions. Both the bounds on cubicity are shown to be tight
upto a factor of $O(\log\log n)$.
\end{abstract}
\bibliographystyle{plain}

\section{Introduction}
Let $\mathcal{F}=\{S_x\subseteq U:x\in V\}$ be a family of subsets of a universe $U$, where $V$ is an index set. The intersection graph $\Omega(\mathcal{F})$ of $\mathcal{F}$ has $V$ as vertex set, and two distinct vertices $x$ and $y$ are adjacent if and only if $S_x\cap S_y\neq\emptyset$. Representations of graphs as the intersection graphs of various geometrical objects is a well studied topic in graph theory. Probably the most well studied class of intersection graphs are the \emph{interval graphs}, where each $S_x$ is a closed interval on the real line. A restricted form of interval graphs, that allow only intervals of unit length, are \emph{indifference graphs}.

A well known
concept in this area of graph theory is the \emph{cubicity}, which
was introduced by F. S. Roberts in 1969  \cite {Roberts}.
This concept generalizes the concept of indifference graphs.
A unit cube in $k$ dimensions ($k$-cube) is a Cartesian product 
$R_1 \times R_2 \times \cdots \times R_k$ where $R_i$ (for $1 \le i \le k$)
is a closed interval of the form $[a_i, a_i + 1]$ on the real line. 
Two $k$-cubes, $(x_1,x_2,\ldots,x_k)$ and $(y_1,y_2,\ldots,y_k)$ are said to have a non-
empty intersection if and only if the intervals $x_i$ and $y_i$ have a non-empty intersection
for $1\leq i\leq k$.
For a  graph $G$, its \emph{cubicity}  is the minimum dimension $k$, such that
$G$ is representable as the intersection graph of $k$-cubes.
We denote the cubicity of a graph $G$ by $\cubi(G)$. 
Note that a $k$-cube representation of $G$ using cubes with unit side length 
is equivalent to a $k$-cube representation where the cubes have side length $c$
for some fixed positive constant $c$. 
The graphs of cubicity $1$ are exactly the class of indifference graphs.
The cubicity of a complete graph is taken to be 0. 
If we require that each vertex correspond to a $k$-dimensional axis-parallel
box $R_1 \times R_2 \times \cdots \times R_k$ where $R_i$ (for $1 \le i \le k$)
is a closed interval of the form $[a_i, b_i]$ on the real line, then
the minimum dimension required to represent $G$ is called its $boxicity$
denoted as $\boxi(G)$. Clearly $\boxi(G) \le \cubi(G)$ for any graph $G$
because cubicity is a stricter notion than boxicity.
It has been shown that deciding whether the cubicity of a given graph is at least 3 is NP-hard \cite{Yan1}.
As for boxicity, it was shown 
by Kratochvil \cite {Kratochvil} 
that deciding  whether the
boxicity of a graph is at most 2  is NP--complete. 

In many algorithmic problems related to graphs, the availability of
certain convenient representations turn out to be extremely useful.
Probably, the most well-known and important examples are the tree decompositions
and path decompositions.
Many NP-hard problems are known to be
polynomial time solvable given a tree(path) decomposition  of the input
graph that has bounded width. Similarly, the representation of
graphs as intersections of ``disks" or ``spheres" lies at the core
of solving problems related to frequency assignments in radio networks,
computing molecular conformations etc. For the  maximum independent set problem
which is hard to approximate  within a factor of $n^{(1/2) - \epsilon}$
for general graphs, a PTAS is known for disk graphs given the disk representation  
\cite{Erl01,Chan01}.
In a similar way,
the availability of cube or box representation in low dimension
makes some well known NP-hard problems like the max-clique problem, 
polynomial time solvable since there are only $O((2n)^k)$ maximal cliques
if the boxicity or cubicity is at most $k$.
Though the maximum independent set problem is hard to approximate within a
factor $n^{(1/2) - \epsilon}$
for general graphs, it is approximable to a $\log n$ factor for boxicity $2$
graphs (the problem is NP-hard  even for boxicity $2$ graphs) given a 
box or cube representation \cite{Agarwal98,Berman2001}.

It is easy to see that the problem of representing graphs using $k$-cubes can be equivalently formulated as
the following geometric embedding problem. Given an undirected unweighted graph
$G=(V,E)$ and a threshold $t$, find an embedding $f : V \rightarrow \mathbb{R}^{k}$
of  the vertices of $G$ into a $k$-dimensional space
(for the minimum possible $k$) such that for any two vertices $u$ and $v$ of $G$,
$||f(u) - f(v)||_{\infty} \le t$ if and only if $u$ and $v$ are adjacent.
The norm $||~~||_{\infty}$ is the $L_{\infty}$ norm. Clearly,
a $k$-cube representation of $G$ yields the required embedding of $G$ in the $k$-dimensional space.
The minimum dimension required to embed $G$ as above under the $L_2$ norm
is called the \emph{sphericity} of $G$.
Refer to \cite{quint2} for applications where 
such an embedding under $L_{\infty}$ norm is argued to be more appropriate than embedding under $L_2$ norm.
The connection between cubicity and sphericity of graphs were studied in \cite{Fishburn,Maehara}.

Roberts \cite{Roberts} showed that for any graph $G$ on $n$ vertices,
$\cubi(G)\leq 2n/3$.
The cube representation of special class of graphs like hypercubes and complete multipartite graphs 
were investigated in \cite{Roberts,Maehara,Quint}.
Similarly, the boxicity of special classes of graphs were studied
by  \cite {Scheiner,Thoma1,CN05}.
An algorithm to compute the box representation in $O(\Delta \ln n)$ dimensions
for any graph $G$ on $n$ vertices and maximum degree $\Delta$  was shown in \cite{tech-rep}.
Researchers have also tried to generalize or extend  the
concept of boxicity in various ways. 
The poset boxicity, 
the rectangle number, grid dimension,
circular dimension   and the boxicity 
of digraphs are some  examples. 

Two recent results about the boxicity of any graph $G$ on $n$ vertices and
having maximum degree $\Delta$ are $\boxi(G)=O(\Delta\ln n)$ \cite{tech-rep}
and $\boxi(G)\leq 2\Delta^2$ \cite{CFNMaxdeg}. Combining these with the result 
$\frac{\cubi(G)}{\boxi(G)}\leq\lceil\log n\rceil$ shown in \cite{CA-06-1},
we get $\cubi(G)=O(\Delta\ln^2 n)$ and $\cubi(G)\leq 2\Delta^2\lceil\log n
\rceil$.
Our result is an improvement over both these bounds on cubicity.
~~~~~~~~~~~~~~~~~~~~~~~~~~~~~~~~~~~~~~~~~~~~~~~~~~~~~~~~~~~~~~~~~~~~~~

\noindent
\emph{Linear arrangement and Bandwidth.} Given an undirected graph $G=(V,E)$
on $n$ vertices,  a \emph{linear arrangement} of the vertices of $G$ is a bijection $L : V \rightarrow \{1,\ldots,n\}$.
The \emph{width} of the linear arrangement $L$ is defined as $\max_{(u,v) \in E} |L(u) - L(v)|$.
The \emph{bandwidth minimization problem} is to compute $L$ with minimum possible width.
The \emph{bandwidth} of $G$ denoted as $b$ is the minimum possible width achieved by any linear
arrangement of $G$. A \emph{bandwidth ordering} of $G$ is a linear arrangement of $V(G)$ with width $b$.

\subsection{Our results} \label{results}
We summarize below the results of this paper.

Let $G$ be a graph on $n$ vertices.
Let $\Delta$ be the maximum degree of $G$ and $b$ its bandwidth.

We first show a randomized algorithm to construct the cube representation of
$G$ in $O(\Delta$ $\ln n)$ dimensions. This randomized construction can be
easily derandomized to obtain a polynomial time deterministic algorithm
that gives a cube representation of $G$ in the same number of dimensions.
We then give a second algorithm that takes as input a linear arrangement
of the vertices of $G$ with width $b$ to construct the $k$-cube representation
of $G$ in $k= O(\Delta \ln b)$ dimensions.
Note that the bandwidth $b$ is at most $n$
and $b$ is much smaller than $n$ for many well-known graph classes.

Note that the second algorithm  to compute the cube representation of a graph $G$ takes as input 
a linear arrangement of $V(G)$. The smaller the width of this arrangement, the lesser
the number of dimensions of the cube representation of $G$ computed by our algorithm.
It is NP-hard to approximate the bandwidth of $G$ within a ratio better than
$k$ for every fixed $k \in \mathbb{N}$ \cite{Ung98}. 
Feige \cite{Feige98} gives a $O(\log^3(n)\sqrt{\log n \log\log n})$ approximation algorithm
to compute the bandwidth (and also the corresponding linear
arrangement) of general graphs using which we obtain polynomial time deterministic or randomized algorithms 
to construct the cube representation of $G$ in $O(\Delta (\ln b + \ln \ln n))$
dimensions, given only $G$.
It should be noted that several algorithms
 with good heuristics that perform
 very well in practice \cite{Turner} are known for bandwidth computation.

We also show that the bounds on cubicity given by both our algorithms
are tight up to a factor of $O(\log\log n)$.
\ignore{
a) $k = O(\Delta (\ln b + \ln \ln n))$ for any graph $G$.
b) $k = O(1)$ if $G$ is a bounded bandwidth graph.
c) $k = O(\Delta)$ if $G$ is either an AT-free graph or a  circular arc graph.
Note that AT-free graphs include well-known graph classes like interval graphs, permutation graphs, co-comparability graphs and trapezoidal graphs.

~~~~~~~~~~~~~~~~~~~~~~~~~~~~~~~~~~~~~~~~~~~~~~~~~~~~~~~~~~~~~~~~~~~~~~

\noindent
\emph{New Structural Results:}
Our algorithmic results yield the following new structural results on the cubicity
of graphs as summarized in the following table.

~~~~~~~~~~~~~~~~~~~~~~~~~~~~~~~~~~~~~~~~~~~~~~~~~~~~~~~~~~~~~~~~~~~~~~~

\begin{tabular}{|l|c|}
\hline
Graph class & Upper bound for cubicity \\\hline\hline
Any graph &  $12 \Delta \ln (2b) + 1$ \\\hline
Any graph  &  $4(\Delta + 1) \ln n$ \\\hline
Any graph & b + 1 \\ \hline
AT-free graphs & $3 \Delta -  1$ \\ \hline
Permutation graphs & $2 \Delta$ \\ \hline
Interval graphs & $\Delta + 1$ \\ \hline
Circular arc graphs & $2\Delta + 1$ \\ \hline
\end{tabular}

Our upper bounds on cubicity for the special graph classes such as AT-free
graphs, permutation graphs and circular arc graphs are tight upto a constant
factor.

We refer the reader to the full version of this paper \cite{CFNfull06} for
the proofs that are omitted due to space constraints.
}

\subsection{Definitions and Notations}

Let $G$ be a simple, finite, undirected graph on $n$ vertices. The vertex
set of $G$ is denoted as $V(G)=\{1,\ldots,n\}$ (or $V$ in short). Let $E(G)$
(or $E$ in short) denote the edge set of $G$. 
Let $G'$ be a graph such that $V(G')=V(G)$. Then $G'$ is a \emph{supergraph}
of $G$ if $E(G)\subseteq E(G')$. We define the \emph{intersection} of two
graphs as follows. If $G_1$ and $G_2$ are two graphs such that $V(G_1)=
V(G_2)$, then the intersection of $G_1$ and $G_2$ denoted as $G=G_1\cap G_2$
is graph with $V(G)=V(G_1)=V(G_2)$ and $E(G)=E(G_1)\cap E(G_2)$. For a vertex
$u\in V(G)$, $N(u)$ denotes the set of neighbours of $u$. The degree of the
vertex $u$ in $G$ is denoted by $d(u)$ and $d(u)=|N(u)|$. Let $\Delta$ denote
the maximum degree of $G$.
Let $b$ denote the bandwidth of $G$.

\subsection{Indifference graph representation}\label{subsecdef}
Let $G=(V,E)$ be a graph and $I_1,\ldots,I_k$ be $k$ indifference graphs such that $V(I_i)=V(G)$ and $E(G)\subseteq E(I_i)$, for $1\leq i\leq k$. If $G=I_1\cap\ldots\cap I_k$, then we say that $I_1,\ldots,I_k$ is an \emph{indifference graph representation} of $G$. The following theorem due to Roberts relates $\cubi(G)$ to the indifference graph representation of $G$.
\begin{theorem}
A graph $G$ has $\cubi(G)\leq k$ if and only if it can be expressed as the intersection of $k$ indifference graphs.
\end{theorem}
All our algorithms compute an indifference graph representation of $G$.
It is straightforward to derive the cube representation of $G$ given its indifference graph representation.
To describe an indifference graph, we define a function $f: V \rightarrow \mathbb{R}$ such that for a vertex $u$,
its closed interval is given by $[f(u), f(u) + l]$, for a fixed constant
$l$ which is assumed to be $1$ unless otherwise specified. Note that even 
if the $l$ value is different for each of the component indifference graphs,
the unit cube representation can be  derived by scaling down all the
intervals of each component indifference graph by the corresponding $l$ value. 

\section{Cube representation in $O(\Delta \ln n)$ dimensions}\label{sectiondlogn}

In this section we describe an algorithm to compute the cube representation of
any graph $G$ on $n$ vertices and maximum degree $\Delta$ in $O(\Delta \ln n)$
dimensions.
\ignore{We achieve this in two steps. First we obtain a simpler algorithm to construct the cube
representation in $O(\Delta \ln n)$ dimensions. Then we improve it to obtain the desired result.

\subsection{Cube representation in \ensuremath{O(\Delta \ln n)} dimensions} \label{sectiondlogn}
}
\begin{definition}
Let $\pi$ be a permutation of the set $\{1,\ldots,n\}$. Let $X\subseteq \{1,
\ldots,n\}$. The projection of $\pi$ onto $X$ denoted as $\pi_X$ is defined
as follows. Let $X=\{u_1,\ldots,u_r\}$ such that $\pi(u_1)<\pi(u_2)<\ldots<
\pi(u_r)$. Then $\pi_X(u_1)=1, \pi_X(u_2)$ $=2,\ldots,\pi_X(u_r)=r$.
\end{definition}

\noindent\textbf{Construction of indifference supergraph given $\pi$:}\\
Let $G(V,E)$ be a simple, undirected graph. Let $\pi$ be a 
permutation on $V$ and let $A$ be a subset of $V$. We define 
$\mathcal{M}(G,\pi,A)$ to be an indifference graph $G'$ constructed as 
follows:

Let $B=V-A$. We now assign intervals of length $n$ to the vertices in $V$. 
The function $f$ defines the left end-points of the intervals (of length 
$n$) mapped to each vertex as follows:\\
$\forall u\in B$, define $f(u)=n+\pi(u)$,\\
$\forall u\in A$ and $N(u)\cap B=\emptyset$, define $f(u)=0$,\\
$\forall u\in A$ and $N(u)\cap B\neq \emptyset$, define $f(u)=\max_{x\in 
N(u)\cap B}\{\pi(x)\}$.

$G'$ is the intersection graph of these intervals. Thus, two vertices $u$ 
and $v$ will have an edge in $G'$ if and only if $|f(u)-f(v)|\leq n$.
Since each vertex is mapped to an interval of length $n$, $G'$ is an 
indifference graph. It can be seen that the vertices in $B$ induce a 
clique in $G'$ as the intervals assigned to each of them contain the point 
$2n$. Similarly, all the vertices in $A$ also induce a clique in $G'$ as 
the intervals mapped to each contain the point $n$.

Now, we show that $G'$ is a supergraph of $G$. To see this, take any edge $(u,v)\in 
E(G)$. If $u$ and $v$ both belong to $A$ or if both belong to $B$, then 
$(u,v)\in E(G')$ as we have observed above. If this is not the case, then 
we can assume without loss of generality that $u\in A$ and $v\in B$. Let 
$t=\max_{x\in N(u)\cap B}\{\pi(x)\}$. Obviously, $t\geq\pi(v)$, since 
$v\in N(u)\cap B$. From the definition of $f$, we have $f(u)=t$ and we 
have $f(v)=n+\pi(v)$. Therefore, $f(v)-f(u)=n+\pi(v)-t$ and since 
$t\geq\pi(v)$, it follows that $f(v)-f(u)\leq n$. This shows that 
$(u,v)\in E(G')$.

We give a randomized algorithm \textbf{RAND} that, given an input graph $G$, outputs an indifference supergraph $G'$ of $G$.

\begin{tabbing}
~~~~~\=~~~~~~~~~~~~\=~~~~~~~~~~~~~~~~~~~~~~~~~\\\kill
\textbf{RAND}\\
\>Input:\> $G$.\\
\>Output:\> $G'$ which is an indifference supergraph of $G$.\\
\textbf{begin}\\
\>Step 1.\>Generate a permutation $\pi$ of $\{1,\ldots,n\}$ uniformly at
random.\\
\>Step 2.\>For each vertex $u\in V$, toss an unbiased coin to decide whether it
\\
\>\>should belong to $A$ or to $B$ (i.e. $\Pr[u\in A]=\Pr[u\in B]=\frac{1}{2}$).\\
\>Step 3.\>Return $G'=\mathcal{M}(G,\pi,A)$.\\
\textbf{end}
\end{tabbing}
\ignore{
\noindent \begin{longtable}{p{0.1in}lp{60mm}}
\multicolumn{3}{l}{\textbf{RAND}}\\
&Input:&$G$.\\
&Output:&$G'$ which is an indifference supergraph of $G$.\\
\multicolumn{3}{l}{\textbf{begin}}\\
&Step 1.&Generate a permutation $\pi$ of $\{1,\ldots,n\}$ uniformly at
random.\\
&Step 2.&For each vertex $u\in V$, toss an unbiased coin to decide whether it
should belong to $A$ or to $B$ (i.e. $\Pr[u\in A]=\Pr[u\in B]=\frac{1}{2}$).\\
&Step 3.&Return $G'=\mathcal{M}(G,\pi,A)$.\\
\multicolumn{3}{l}{\textbf{end}}
\end{longtable}
}

\begin{lemma} \label{rand-basic-lemma}
Let $e=(u,v)\notin E(G)$. Let $G'$ be the graph returned by \textbf{RAND}($G$). Then,
\begin{eqnarray*}
\Pr[e\in E(G')]&\leq&\frac{1}{2}+\frac{1}{4}\left(\frac{d(u)}{d(u)+1}+
\frac{d(v)}{d(v)+1}\right)\\
&\leq&\frac{2\Delta+1}{2\Delta+2}
\end{eqnarray*}
where $\Delta$ is the maximum degree of $G$.
\end{lemma}
\begin{proof}
Let $\pi$ be the permutation and $\{A,B\}$ be the partition of $V$ generated
randomly by \textbf{RAND}($G$).
An edge $e=(u,v) \notin E(G)$ will be present in $G'$ if and only if one of the following cases occur:
\begin{enumerate}
\item{Both $u,v\in A$ or both $u,v\in B$}
\item{$u\in A, v\in B$ and $\max_{x\in N(u)\cap B}\pi(x)>\pi(v)$}
\item{$u\in B, v\in A$ and $\max_{x\in N(v)\cap B}\pi(x)>\pi(u)$}
\end{enumerate}
Let $P_1$ denote the probability of situation 1 to occur, $P_2$ that of situation 2 and $P_3$ that of situation 3. Since all the three cases are mutually exclusive, $\Pr[e\in E(G')]=P_1+P_2+P_3$. It can be easily seen that $P_1=\Pr[u,v\in A]+\Pr[u,v\in B]=\frac{1}{4}+\frac{1}{4}=\frac{1}{2}$. $P_2$ and $P_3$ can be calculated as follows: 
\begin{displaymath}
P_2=\Pr[u\in A \wedge v\in B \wedge \max_{x\in N(u)\cap B}\pi(x)>\pi(v)]
\end{displaymath}
Note that creating the random permutation and tossing the coins are two different experiments independent of each other. Moreover, the coin toss for each vertex is an experiment independent of all other coin tosses.
Thus, the events $u\in A$, $v\in B$ and $max_{x\in N(u)\cap B}\pi(x)>\pi(v)$ are all independent of each other. Therefore,
\begin{align*}
P_2=\Pr[u\in A]&\times\Pr[v\in B]\times \\
&\Pr[\max_{x\in N(u)\cap B}\pi(x)>\pi(v)]
\end{align*}
Now, $\Pr[\max_{x\in N(u)\cap B}\pi(x)>\pi(v)] \leq$
$\Pr[\max_{x\in N(u)}\pi(x)>\pi(v)]=p$ (say). Let $X=\{v\}\cup N(u)$ and let $\pi_X$ be the projection of $\pi$ onto $X$. Then $p$ is the probability that the condition $\pi_X(v) \neq |X|$ is satisfied. Since $\pi_X$ can be any permutation of $|X|=d(u)+1$ elements with equal probability $\frac{1}{(d(u)+1)!}$ and the number of permutations which satisfy our condition is $d(u)!d(u)$,  $p=\frac{d(u)!d(u)}{(d(u)+1)!} = \frac{d(u)}{d(u)+1}$. Therefore, $\Pr[\max_{x\in N(u)\cap B}\pi(x)>\pi(v)] \leq \frac{d(u)}{d(u)+1}$. It can be easily seen that $\Pr[u\in A]=\frac{1}{2}$ and $\Pr[v\in B]=\frac{1}{2}$. Thus,
\begin{displaymath}
P_2 \leq \frac{1}{2}\times\frac{1}{2}\times\frac{d(u)}{d(u)+1} = \frac{1}{4}\left(\frac{d(u)}{d(u)+1}\right)
\end{displaymath}
Using similar arguments,
\begin{displaymath}
P_3 \leq \frac{1}{4}\left(\frac{d(v)}{d(v)+1}\right)
\end{displaymath}
Thus,
\begin{align*}
\Pr[e\in E(G')]&=P_1+P_2+P_3 \\
&\leq\frac{1}{2}+\frac{1}{4}\left(\frac{d(u)}{d(u)+1}+\frac{d(v)}{d(v)+1}\right)
\end{align*}
Hence the lemma.
\end{proof}

\begin{theorem}\label{cubdelta}
Given a simple, undirected graph $G$ on $n$ vertices with maximum degree $\Delta$, $\cubi(G)\leq \lceil 4(\Delta +1)\ln n\rceil$.
\end{theorem}
\begin{proof}
Let us invoke \textbf{RAND}($G$) $k$ times so that we obtain $k$ indifference supergraphs of $G$ which we will call $G_1',G_2',\ldots,G_k'$. Let $G''=G_1'\cap G_2'\cap\ldots\cap G_k'$. Obviously, $G''$ is a supergraph of $G$. If $G''=G$, then we have obtained $k$ indifference graphs whose intersection gives $G$, which in turn means that $\cubi(G)\leq k$. The $k$ indifference graphs can be seen as an indifference graph representation of $G$. We now estimate an upper bound for the value of $k$ so that $G''=G$.

Let $(u,v)\notin E(G)$.
\begin{align*}
 \Pr[(u,v)\in&E(G'')]=\Pr\left[\bigwedge_{1\leq i\leq k}(u,v)\in E(G_i')\right]\\
 & \leq\left(\frac{2\Delta+1}{2\Delta+2}\right)^k\mbox{(From lemma 1)}
\end{align*}
\begin{eqnarray*}
\Pr[G''\neq G]&=&\Pr\left[\bigvee_{(u,v)\notin E(G)}(u,v)\in E(G'')\right]\\
& \leq&  \frac{n^2}{2}\left(\frac{2\Delta+1}{2\Delta+2}\right)^k\\
& = & \frac{n^2}{2}\left(1-\frac{1}{2(\Delta+1)}\right)^k\\
& \leq& \frac{n^2}{2}\times e^{-\frac{k}{2(\Delta+1)}}
\end{eqnarray*}\\
Choosing $k=4(\Delta+1)\ln n$, we get,
\begin{displaymath}
\Pr[G''\neq G]\leq \frac{1}{2}
\end{displaymath}
Therefore, if we invoke \textbf{RAND} $\lceil4(\Delta+1)\ln n\rceil$ times, there is a non-zero probability that we obtain an indifference graph representation of $G$. Thus, $\cubi(G)\leq \lceil4(\Delta+1)\ln n\rceil$.
\end{proof}

\begin{theorem} \label{thmhp}
Given a graph $G$ on $n$ vertices with maximum degree $\Delta$. Let
$G_1,$ $G_2,\ldots,G_k$ be $k$ indifference supergraphs of $G$ generated by $k$
invocations of \textbf{RAND}($G$) and let $G''=G_1'\cap G_2'\cap\ldots\cap
G_k'$. Then, for $k\geq 6(\Delta+1)\ln n$, $G''=G$ with high probability.
\end{theorem}
\begin{proof}
Choosing $k=6(\Delta+1)\ln n$ in the final step of proof of theorem \ref{cubdelta}, we get,
\begin{displaymath}
\Pr[G''\neq G]\leq \frac{1}{2n}
\end{displaymath}
Thus, if $k\geq 6(\Delta+1)\ln n$, $G''=G$ with high probability.
\end{proof}

\begin{theorem} \label{thmalg}
Given a graph $G$ with $n$ vertices, $m$ edges and maximum degree $\Delta$, with high probability, its cube representation in $\lceil6(\Delta+1)\ln n\rceil$ dimensions can be generated in $O(\Delta(m+n)\ln n)$ time.
\end{theorem}
\begin{proof}
We assume that a random permutation $\pi$ on $n$ vertices can be computed in $O(n)$ time and that a random coin toss for each vertex takes only $O(1)$ time. We take $n$ steps to assign intervals to the $n$ vertices. Suppose in a given step, we are attempting to assign an interval to vertex $u$. If $u\in B$, then we can assign the interval $[n+\pi(u),2n+\pi(u)]$ to it in constant time. If $u\in A$, We look at each neighbour of the vertex $u$ in order to find out a neighbour $v\in B$ such that $\pi(v)=\max_{x\in N(u)\cap B}\pi(x)$ and assign the interval $[\pi(v),n+\pi(v)]$ to $u$. It is obvious that determining this neighbour $v$ will take just $O(d(u))$ time. Since the number of edges in the graph $m=\frac{1}{2}\Sigma_{u\in V}d(u)$, one invocation of \textbf{RAND} needs only $O(m+n)$ time. Since we need to invoke \textbf{RAND} $O(\Delta\ln n)$ times (see the proof of Theorem \ref{cubdelta}), the overall algorithm that generates the cube representation in $6(\Delta+1)\ln n$ dimensions runs in $O(\Delta(m+n)\ln n)$ time.
\end{proof}


\noindent\textbf{Derandomization:} The above algorithm can be derandomized by adapting the 
techniques used in \cite{tech-rep} to obtain a deterministic polynomial time algorithm \textbf{DET} with the
same performance guarantee on the number of dimensions for the cube representation.
Let $t=\lceil 4(\Delta$ $+1)\ln n\rceil$. Given $G$, \textbf{DET} selects $t$ permutations
$\pi_1,\ldots,\pi_t$ and $t$ subsets $A_1,\ldots,A_t$ of $V$ in such a way that
the indifference graphs $\{\mathcal{M}(G,\pi_i,A_i)~|~ 1\leq i\leq t\}$ form an
indifference graph representation of $G$.

\section{Improving to  $O(\Delta \ln b)$ dimensions}

In this section we show an algorithm \textbf{DETBAND} to construct the cube
representation
of $G=(V,E)$ in $O(\Delta \ln b)$ dimensions given a linear arrangement
$\mathcal{A}$ of $V(G)$ with width $b$. The \textbf{DETBAND} algorithm internally invokes
the \textbf{DET} algorithm (see the derandomization part of Section \ref{sectiondlogn}).
Let the linear arrangement $\mathcal{A}$ be $v_1, \ldots, v_n$.
For ease of presentation, assume that $n$ is a multiple of $b$. 
Define a partition $B_0,\ldots, B_{k-1}$ of $V$ where $k=n/b$,
where $B_j = \{v_{jb + 1},\ldots, v_{jb + b}\}$. Let $H_i$ for $0 \le i \le
k-2$ be the induced subgraph of $G$ on the vertex set $B_i \cup B_{i+1}$.
Since for any $i$, $|V(H_i)|=2b$, we have $\cubi(H_i)\leq \lceil
4(\Delta+1)\ln (2b)\rceil=t$ (say). Let $H_i^1,\ldots,H_i^t$ be the indifference
graph representation for $H_i$ produced by \textbf{DET} when given $H_i$ as the
input. Let $g_i^1,\ldots,g_i^t$ be their corresponding unit interval
representations.

Define, for $0\leq i\leq 2$, the graph $G_i$ with $V(G_i)=V$ as the
intersection of $t$ indifference graphs $I_{i,1},\ldots,I_{i,t}$.
Let $f_{i,j}$ be the unit interval representation for $I_{i,j}$.
For each vertex $u\in V$, define $f_{i,j}(u)$ as follows:\\
If $u\in V(H_s)$ such that $s\in \{i,i+3,i+6,\ldots\}$, then define
$f_{i,j}(u)=g_s^j(u)$.\\
Otherwise\ignore{(i.e., if $u\in B_{i+2}\cup B_{i+5}\cup\cdots$)}, define
$f_{i,j}(u)=n$.

The indifference graph $I_0$ is constructed by assigning to each vertex
in $B_i$ the interval $[in,(i+1)n]$, for $0\leq i\leq k-1$.

We prove that $G=I_0\cap G_0\cap G_1\cap G_2$ and thereby show that
$\cubi(G)\leq 12(\Delta+1)\lceil\ln(2b)\rceil +1$ or $\cubi(G)=O(\Delta\ln b)$.

\begin{definition}
Let $I_1$ and $I_2$ be two indifference graphs on disjoint sets of vertices
$V_1$ and $V_2$ respectively.
Let $f_1$ and $f_2$ be their corresponding unit interval representations.
We say that a unit interval graph representation $f: V_1 \cup V_2 \rightarrow
\mathbb{R}$ of $I_1 \cup I_2$ is a union of $f_1$ and $f_2$ if 
$f(u) = f_1(u)$ if $u \in V_1$ and $f(u) = f_2(u)$ if $u \in V_2$.
\end{definition}

\ignore{
we overlap the unit interval representations
of $H_0^j,H_3^j,H_6^j,\ldots$ on the region $[0,3n]$ on the real line and
for all the vertices in $V-(V(H_0)\cup V(H_3)\cup\ldots)=B_2\cup B_5\cup
\ldots$, we assign the interval $[n,2n]$ to get another indifference
graph $I_{0,j}$.
The idea here is to construct $3t+1$ indifference graphs $\{I_0\}\cup
\{I_{0,j},I_{1,j},I_{2,j}~|~1\leq j\leq t\}$. 
a unit interval representation for $H_i$ for
each $i$ using $\lceil 4(\Delta+1) \ln (2b)\rceil$ dimensions by invoking
\textbf{DETBAND} on $H_i$ (Note that $|H_i|=2b$). 
}
Let $t=\lceil 4(\Delta+1)\ln (2b)\rceil$.\vspace{0.1in}

\ignore{
\noindent\begin{longtable}{p{0.1in}p{0.1in}p{70mm}}
\multicolumn{3}{l}{\textbf{DETBAND}}\\
&&\\
&\multicolumn{2}{l}{Input: $G, \mathcal{A}$}\\
&\multicolumn{2}{l}{\parbox{73mm}{Output: Representation of $G$ using $3t + 1$
indifference graphs}}\\
&&\\
\multicolumn{3}{l}{\textbf{begin}}\\
&\multicolumn{2}{l}{(The length of each interval is $n$)}\\
&&\\
&\multicolumn{2}{l}{\parbox{73mm}{Construct $I_0$: for each $i$ and for each node $v \in B_i$,
$f_0(v) = i\cdot n$.}}\\
&&\\
&\multicolumn{2}{l}{\parbox{73mm}{Construction of $I_{i,j}$, $0 \le i \le 2$ and $1 \le j \le
t$:}}\vspace{0.1in}\\
&&Invoke \textbf{DET} on each induced subgraph in $\mathcal{H}
= \{H_{3r + i}: r=0,1,\ldots\}$.\vspace{0.1in}\\
&&Let $H_k^1,\ldots,H_k^t$ be the indifference graphs output by \textbf{DET} for $H_k$.\vspace{0.1in}\\
&&Let $S = V-\bigcup_{H\in\mathcal{H}}V(H)$.\vspace{0.1in}\\
&&Let $f_S : S \rightarrow \mathbb{R}$ be defined as $f_S(v) = n$ for all $v
\in S$.\vspace{0.1in}\\
&&Define $f_{i,j}$ as the \emph{union} of $f_S$ and the $f$ functions
corresponding to each graph in $\{H_{3r +i}^j : r = 0,1,2,\ldots\}$.\vspace{0.1in}\\
\multicolumn{3}{l}{\textbf{end}}
\end{longtable}
}

\begin{tabbing}
~~~~~\=~~~~~\=~~~~~~~~~~~~~~~~~~~~\=~~~~~\\\kill
\textbf{DETBAND}\\
\>Input: $G, \mathcal{A}$.\\
\>Output: Representation of $G$ using $3t + 1$ indifference graphs.\\
\textbf{begin}\\
\>(The length of each interval is $n$)\\\\
\> Construct $I_0$: for each $i$ and for each node $v \in B_i$, $f_0(v) = i\cdot n$.\\\\
\> Construction of $I_{i,j}$, $0 \le i \le 2$ and $1 \le j \le t$:\\
\>\>Invoke \textbf{DET} on each induced subgraph in $\mathcal{H} = \{H_{3r + i}: r=0,1,\ldots\}$.\\
\>\>Let $H_k^1,\ldots,H_k^t$ be the indifference graphs output by \textbf{DET} for $H_k$.\\
\>\>Let $S = V-\bigcup_{H\in\mathcal{H}}V(H)$.\\
\>\>Let $f_S : S \rightarrow \mathbb{R}$ be defined as $f_S(v) = n$ for all $v \in S$.\\
\>\>Define $f_{i,j}$ as the \emph{union} of $f_S$ and the $f$ functions corresponding to \\
\>\>\>each graph in $\{H_{3r +i}^j : r = 0,1,2,\ldots\}$.\\
\textbf{end}\\
\end{tabbing}

\begin{theorem}\label{detband}
\textbf{DETBAND} constructs the cube representation of $G$ in $12(\Delta+1)\lceil\ln(2b)
\rceil+1$ dimensions in polynomial time.
\end{theorem}
\begin{proof}
Let $t=\lceil 4(\Delta+1)\ln(2b)\rceil$.
\pagebreak
\begin{claim}
$I_0$ is a supergraph of $G$.
\end{claim}
\begin{proof}
Consider an edge $(v_x,v_y)\in E(G)$ (assume $x<y$). If $B_m$ is the block containing $v_x$, then $v_y$ is contained in either $B_m$ or $B_{m+1}$ since $y-x\leq b$ and each block contains $b$ vertices. Thus, $f_0(v_x)=mn$ and $f_0(v_y)=mn$ or $mn+n$. In either case, there is an overlap between $f_0(v_x)$ and $f_0(v_y)$ at the point $mn$ and therefore, $(v_x,v_y)\in E(I_0)$.
\end{proof}
\begin{claim}
$I_{i,j}$, for $0\leq i\leq 2$ and $1\leq j\leq t$, is a supergraph of $G$.
\end{claim}
\begin{proof}
Consider an edge $(v_x,v_y)\in E(G)$ (assume $x<y$). Let $B_m$ be the block that contains $v_x$. From our earlier observation, $v_y$ is either in $B_m$ or in $B_{m+1}$.

If $v_x,v_y\in V(H_p)$, where $p=3r+i$ for some $r\geq 0$, then by definition of $f_{i,j}$, $f_{i,j}(v_x)$ and $f_{i,j}(v_y)$ correspond to the intervals assigned to them in the interval representation of the indifference graph $H_p^j$. Since $(v_x,v_y)\in E(H_p)$ and $H_p\subseteq H_p^j$, the intervals $f_{i,j}(v_x)$ and $f_{i,j}(v_y)$ overlap and therefore $(v_x,v_y)\in E(I_{i,j})$.

Now, if $m=3r+i$, for some $r\geq 0$, then $v_x,v_y\in H_m$. Therefore, as detailed in the previous paragraph, it follows that $(v_x,v_y)\in E(I_{i,j})$.

If $m=3r+i+1$, for some $r\geq 0$, then either $v_x,v_y\in V(H_{m-1})$ or $v_x\in V(H_{m-1})$ and $v_y\in S$. In the first case, the earlier argument can be applied again to obtain the result that $(v_x,v_y)\in E(I_{i,j})$. Now, if $v_x\in V(H_{m-1})$ and $v_y\in S$, we have $m-1=3r+i$ and therefore by definition of $f_{i,j}$, $f_{i,j}(v_x)$ is the interval mapped to $v_x$ in the interval representation of the indifference graph $H_{m-1}^j$. From the construction of \textbf{DET}, it is clear that $0\leq f_{i,j}(v_x)\leq 2n$ (see the derandomization part of section \ref{sectiondlogn}). Also, we have $f_{i,j}(v_y)=f_S(v_y)=n$. It can be seen that $|f_{i,j}(v_x)-f_{i,j}(v_y)|\leq n$ and therefore $(v_x,v_y)\in E(I_{i,j})$.

Similarly, if $m=3r+i+2$, for some $r\geq 0$, then $v_x\in S$ and $v_y$ is contained either in $S$ or in $V(H_{m+1})$. It can be shown using arguments similar
to the ones used in the preceding paragraph that $(v_x,v_y)\in E(I_{i,j})$.
\ignore{
If $v_y\in S$, then $f_{i,j}(v_x)=f_{i,j}(v_y)=n$, and therefore $(v_x,v_y)\in E(I_{i,j})$. If $v_y\in V(H_{m+1})$, we have $m+1=3(r+1)+i$, and therefore by the construction of \textbf{DET}, $0\leq f_{i,j}(v_y)\leq 2n$. We know that as $v_x\in S$, $f_{i,j}(v_x)=n$. Therefore, $|f_{i,j}(v_x)-f_{i,j}(v_y)|\leq n$ which implies that $(v_x,v_y)\in E(I_{i,j})$.
}

This completes the proof that $G\subseteq I_{i,j}$, for $0\leq i\leq 2$, $1\leq j\leq t$.
\end{proof}
\begin{claim}
The indifference graphs $I_{i,j}$, for $0\leq i\leq 2$ and $1\leq j\leq t$, along with $I_0$ constitute a valid indifference graph representation of $G$.
\end{claim}
\begin{proof}
We have to show that given any edge $(v_x,v_y)\not\in E(G)$, there is at least one graph among the $3t+1$ indifference graphs generated by \textbf{DETBAND} that does not contain that edge.

Assume that $x<y$. Let $B_m$ and $B_l$ be the blocks containing $v_x$ and
$v_y$ respectively. If $l-m>1$ then $f_0(v_y)-f_0(v_x)=(l-m)n>n$. Therefore,
$(v_x,v_y)\not\in E(I_0)$. Now we consider the case when $l-m\leq 1$. Consider
the set of indifference graphs $\mathcal{I}=\{H_m^j ~|~ 1\leq j\leq t\}$ that
is generated by \textbf{DET} when given $H_m$ as input. We know that $(v_x,v_y)\not\in
E(H_m)$ because $H_m$ is an induced subgraph of $G$ containing the vertices
$v_x$ and $v_y$. Since $\mathcal{I}$ is a valid indifference graph 
representation of $H_m$, at least one of the graphs in $\mathcal{I}$, say 
$H_m^p$, should not contain the edge $(v_x,v_y)$. Let $g$ be the unit interval
representation function corresponding to $H_m^p$ as output by \textbf{DET}. Since
$(v_x,v_y)\not\in E(H_m^p)$, $|g(v_x)-g(v_y)|>n$. Let $i=m \mod 3$. Thus,
$m=3r+i$, for some $r\geq 0$. Now, since $f_{i,p}$ is defined as the union of
the unit interval representation functions of all the graphs in the set
$\{H_{3r +i}^p : r = 0,1,2,\ldots\}$ which contains $H_m^p$, $f_{i,p}(v_x)=
g(v_x)$ and $f_{i,p}(v_y)=g(v_y)$ which implies that $|f_{i,p}(v_x)-f_{i,p}
(v_y)|>n$. Therefore, $(v_x,v_y)\not\in E(I_{i,p})$.
\end{proof}

Thus, \textbf{DETBAND} generates a valid indifference graph representation of $G$ using
$3t+1\leq 12(\Delta+1)\lceil\ln(2b)\rceil$ indifference graphs. Since \textbf{DET} runs
in polynomial time and there are only polynomial number of invocations of \textbf{DET}, 
the procedure \textbf{DETBAND} runs in polynomial time.
\end{proof}

{\textbf{Tight example:}} Consider the case when $G$ is a complete binary 
tree of height $d=\log n$. Using the results shown in \cite{CMO}, we can see 
that $\cubi(G)\geq \frac{d}{\log 2d}=\frac{\log n}{c_1+\log\log n}$ where 
$c_1$ is a constant. Therefore, $\cubi(G)=\Omega(\frac{\log n}{\log\log n})$. 
From theorem \ref{cubdelta}, $\cubi(G)\leq 4(\Delta+1)\ln n=16\ln n=c_2\log n$,
where $c_2$ is a constant. Therefore, the upper bound provided by theorem
\ref{cubdelta} is tight up to a factor of $O(\log\log n)$. Since the bandwidth
of the complete binary tree on $n$ vertices is $\Theta(\frac{n}{\log n})$ as
shown in \cite{heck}, the
$O(\Delta\ln b)$ bound on cubicity is also tight up to a factor of $O(\log\log
n)$.


\begin{thebibliography}{10}

\bibitem{Chan01}
P.~Afshani and T.~Chan.
\newblock Approximation algorithms for maximum cliques in 3d unit-disk graphs.
\newblock In {\em Proc. 17th Canadian Conference on Computational Geometry
  (CCCG)}, pages 6--9, 2005.

\bibitem{Agarwal98}
P.~K. Agarwal, M.~van Kreveld, and S.~Suri.
\newblock Label placement by maximum independent set in rectangles.
\newblock {\em Comput. Geom. Theory Appl.}, 11:209--218, 1998.

\bibitem{Berman2001}
P.~Berman, B.~DasGupta, S.~Muthukrishnan, and S.~Ramaswami.
\newblock Efficient approximation algorithms for tiling and packing problems
  with rectangles.
\newblock {\em J. Algorithms}, 41:443--470, 2001.

\bibitem{tech-rep}
L.~Sunil Chandran, Mathew~C. Francis, and Naveen Sivadasan.
\newblock Geometric representation of graphs in low dimension.
\newblock To appear in Algorithmica, available at
  http://arxiv.org/abs/cs.DM/0605013.

\bibitem{CFNMaxdeg}
L.~Sunil Chandran, Mathew~C. Francis, and Naveen Sivadasan.
\newblock Boxicity and maximum degree.
\newblock {\em Journal of Combinatorial Theory, Series B}, 98(2):443--445,
  March 2008.

\bibitem{CMO}
L.~Sunil Chandran, C.~Mannino, and G.~Orialo.
\newblock On the cubicity of certain graphs.
\newblock {\em Information Processing Letters}, 94:113--118, 2005.

\bibitem{CA-06-1}
L.~Sunil Chandran and K.~Ashik Mathew.
\newblock An upper bound for cubicity in terms of boxicity.
\newblock Submitted, 2006.

\bibitem{CN05}
L.~Sunil Chandran and Naveen Sivadasan.
\newblock Boxicity and treewidth.
\newblock {\em Journal of Combinatorial Theory, Series B}, 97(5):733--744,
  September 2007.

\bibitem{Erl01}
T.~Erlebach, K.~Jansen, and E.~Seidel.
\newblock Polynomial-time approximation schemes for geometric intersection
  graphs.
\newblock {\em SIAM Journal on Computing}, 34(6):1302--1323, 2005.

\bibitem{Feige98}
Uriel Feige.
\newblock Approximating the bandwidth via volume respecting embeddings.
\newblock In {\em Prceedings of the Thirtieth Annual ACM Symposium on Theory of
  Computing}, pages 90--99. ACM Press, 1998.

\bibitem{Fishburn}
Peter~C. Fishburn.
\newblock On the sphericity and cubicity of graphs.
\newblock {\em Journal of Combinatorial Theory, Series B}, 35(3):309--318,
  December 1983.

\bibitem{heck}
Ralf Heckmann, Ralf Klasing, Burkhard Monien, and Walter Unger.
\newblock Optimal embedding of complete binary trees into lines and grids.
\newblock {\em Journal of Parallel and Distributed Computing}, 49(1):40--56,
  1998.

\bibitem{Kratochvil}
J.~Kratochvil.
\newblock A special planar satisfiability problem and a consequence of its
  {NP}--completeness.
\newblock {\em Discrete Applied Mathematics}, 52:233--252, 1994.

\bibitem{Maehara}
H.~Maehara.
\newblock Sphericity exceeds cubicity for almost all complete bipartite graphs.
\newblock {\em Journal of Combinatorial Theory, Series B}, 40(2):231--235,
  April 1986.

\bibitem{quint2}
T.S. Michael and Thomas Quint.
\newblock Sphere of influence graphs and the $l_{\infty}$-metric.
\newblock {\em Discrete Applied Mathematics}, 127:447--460, 2003.

\bibitem{Quint}
T.S. Michael and Thomas Quint.
\newblock Sphericity, cubicity, and edge clique covers of graphs.
\newblock {\em Discrete Applied Mathematics}, 154(8):1309--1313, May 2006.

\bibitem{Roberts}
F.~S. Roberts.
\newblock {\em Recent Progresses in Combinatorics}, chapter On the boxicity and
  cubicity of a graph, pages 301--310.
\newblock Academic Press, New York, 1969.

\bibitem{Scheiner}
E.~R. Scheinerman.
\newblock Intersection classes and multiple intersection parameters.
\newblock Ph. D thesis, Princeton University, 1984.

\bibitem{Thoma1}
C.~Thomassen.
\newblock Interval representations of planar graphs.
\newblock {\em Journal of Combinatorial Theory, Series B}, 40:9--20, 1986.

\bibitem{Turner}
J.~Turner.
\newblock On the probable performance of heuristics for bandwidth minimization.
\newblock {\em {SIAM} journal on computing}, 15:561--580, 1986.

\bibitem{Ung98}
W.~Unger.
\newblock The complexity of the approximation of the bandwidth problem.
\newblock In {\em Proceedings of the 39th IEEE Annual Symposium on Foundations
  of Computer Science}, pages 82--91, November 1998.

\bibitem{Yan1}
Mihalis Yannakakis.
\newblock The complexity of the partial order dimension problem.
\newblock {\em {SIAM} Journal on Algebraic Discrete Methods}, 3:351--358, 1982.

\end{thebibliography}
\end{document}